\documentclass[times, sort&compress, 3p]{elsarticle}
\journal{Journal of Multivariate Analysis}

\usepackage{amsmath,amsfonts,amssymb,amsthm,booktabs,graphicx,url}

\pdfoutput=1
\usepackage[english]{babel}
\usepackage{enumerate}
\usepackage{mathrsfs}
\usepackage{amsbsy}

\usepackage{natbib}
\usepackage{float}
\usepackage{accents} 

\newtheorem{theorem}{Theorem}
\newtheorem{proposition}{Proposition}

\newtheorem{lemma}{Lemma}
\newtheorem{state}{Statement}

\usepackage[mathscr]{euscript}
\usepackage{enumitem}
\usepackage[table]{xcolor}
\usepackage[flushleft]{threeparttable}

\usepackage{relsize}
\usepackage{algorithm}
\usepackage{algpseudocode}
\usepackage{algorithmicx}
\usepackage{array}
\usepackage{bookmark}

\def\@font@info#1{%
}%
\def\@font@warning#1{%
}%
\def\@pr@videpackage[#1]{%
  \expandafter\xdef\csname ver@\@currname.\@currext\endcsname{#1}%
  \ifx\@currext\@clsextension
    \typeout{Document Class: \@gtempa\space#1}%
  \else
  \fi}

\makeatletter
\newcommand*{\trasp}{%
  {\mathpalette\@trasp{}}%
}
\newcommand*{\@trasp}[2]{%
  ^{\raisebox{\depth}{$\m@th#1\scriptstyle{\top}$}}
}
\makeatother

\makeatletter
\newcommand*{\traspj}{%
  {\mathpalette\@traspj{}}%
}
\newcommand*{\@traspj}[2]{%
  ^{\raisebox{\depth}{$\m@th#1\scriptstyle{\top}$}}_j
}
\makeatother

\newcommand{\trasps}[1]{^{\raisebox{\depth}{$\scriptstyle{\top}$}}_{#1}} 
\newcommand{\traspd}[1]{^{\raisebox{-1.0mm}{$\scriptstyle{\top}$}}_{#1}} 

\newcommand{\traspsM}[1]{^{\raisebox{\depth}{$\scriptstyle{\top}$}}_{{#1}_{[j - 1]}}} 
\newcommand{\traspsMb}[1]{^{\raisebox{\depth}{$\scriptstyle{\top}$}}_{\mathbf{#1}_{[j - 1]}}} 

\newcommand{\matinv}{^{\raisebox{+1.0mm}{$\scriptstyle{{-1}}$}}} 
\newcommand{\matinvj}[1]{^{\raisebox{-1.0mm}{$\scriptstyle{{-1}}$}}_{#1}} 
\newcommand{\matginv}{^{\raisebox{+1.0mm}{$\scriptstyle{{+}}$}}} 

\newcommand{\cspace}{\mathscr{C}}

\newcommand{\ju}{{{j_1}}}

\newcommand{\jind}[1]{{{j_#1}}}

\renewcommand{\Re}{\mathbb{R}}

\newcommand{\bhR}{{\widehat{\mathbf{R}}}}

\newcommand{\hr}{\widehat{\mathbf{r}}}
\newcommand{\hrj}{{\widehat{\mathbf{r}}_j}}
\newcommand{\hrjT}{{\widehat{\mathbf{r}}\trasps{j}}}
\newcommand{\hrju}{{\widehat{\mathbf{r}}_{j_1}}}
\newcommand{\hrjuT}{{\widehat{\mathbf{r}}\trasps{{j_1}}}}

\newcommand{\hwj}{{\widehat{\mathbf{w}}_j}}
\newcommand{\hwju}{{\widehat{\mathbf{w}}_{j_1}}}

\newcommand{\ba}{\mathbf{a}}

\newcommand{\bd}{{\mathbf{d}}}
\newcommand{\bb}{{\mathbf{b}}}

\newcommand{\bt}{{\mathbf{t}}}
\newcommand{\bp}{{\mathbf{p}}}
\newcommand{\bu}{{\mathbf{u}}}
\newcommand{\buj}{{\mathbf{u}_j}}
\newcommand{\bujT}{{\mathbf{u}\trasps{j}}}

\newcommand{\bx}{{\mathbf{x}}}
\newcommand{\by}{{\mathbf{y}}}
\newcommand{\br}{{\mathbf{r}}}

\newcommand{\bz}{{\mathbf{z}}}

\newcommand{\bv}{{\mathbf{v}}}
\newcommand{\bX}{{\mathbf{X}}}
\newcommand{\bXT}{{\mathbf{X}\trasp}}

\newcommand{\bA}{{\mathbf{A}}}

\newcommand{\bC}{{\mathbf{C}}}

\newcommand{\bG}{{\mathbf{G}}}
\newcommand{\bE}{{\mathbf{E}}}

\newcommand{\bT}{{\mathbf{T}}}
\newcommand{\bM}{{\mathbf{M}}}
\newcommand{\bP}{{\mathbf{P}}}
\newcommand{\bR}{{\mathbf{R}}}
\newcommand{\bw}{{\mathbf{w}}}
\newcommand{\bS}{{\mathbf{S}}}

\newcommand{\bU}{{\mathbf{U}}}
\newcommand{\bZ}{{\mathbf{Z}}}
\newcommand{\bY}{{\mathbf{Y}}}
\newcommand{\bW}{{\mathbf{W}}}
\newcommand{\bI}{{\mathbf{I}}}
\newcommand{\bJ}{{\mathbf{J}}}
\newcommand{\bQ}{{\mathbf{Q}}}

\newcommand{\bTd}{{\mathbf{T}_{[d]}}}
\newcommand{\bTdT}{{\mathbf{T}\traspd{[d]}}}

\newcommand{\bAd}{{\mathbf{A}_{[d]}}}
\newcommand{\bAdT}{{\mathbf{A}\traspd{[d]}}}

\newcommand{\bXj}{{\mathbf{X}_j}}
\newcommand{\bXjT}{{\mathbf{X}\trasps{j}}}
\newcommand{\bQj}{{\mathbf{Q}_j}}
\newcommand{\bQjT}{{\mathbf{Q}\trasps{j}}}
\newcommand{\bRj}{{\mathbf{R}_j}}

\newcommand{\brji}{{\mathbf{r}_{j_i}}}

\newcommand{\bMj}{{\mathbf{M}_{j}}}

\newcommand{\bHj}{{\mathbf{H}_j}}
\newcommand{\bHjT}{{\mathbf{H}\trasps{j}}}

\newcommand{\bTjmu}{{\mathbf{T}_{[j-1]}}}

\newcommand{\bYjmu}{{\mathbf{Y}_{[j-1]}}}
\newcommand{\bYjmuT}{{\mathbf{Y}\trasps{[j-1]}}}

\newcommand{\btj}{{\mathbf{t}_j}}
\newcommand{\btjT}{{\mathbf{t}\trasps{j}}}
\newcommand{\baj}{{\mathbf{a}_j}}
\newcommand{\bajT}{{\mathbf{a}\trasps{j}}}
\newcommand{\bvj}{{\mathbf{v}_j}}

\newcommand{\bzj}{{\mathbf{z}_j}}

\newcommand{\bbj}{{\mathbf{b}_j}}

\newcommand{\byj}{{\mathbf{y}_j}}

\newcommand{\brju}{{\mathbf{r}_{j_1}}}
\newcommand{\brjuT}{{\mathbf{r}\trasps{j_1}}}
\newcommand{\bwju}{{\mathbf{w}_{j_1}}}

\newcommand{\dwju}{{\overset{\boldsymbol{.}}{\mathbf{w}}_{j_1}}}
\newcommand{\cwju}{{\overset{\boldsymbol{\sim}}{\mathbf{w}}_{j_1}}}
\newcommand{\dwjuT}{{\overset{\boldsymbol{.}}{\mathbf{w}}\traspd{j_1}}}
\newcommand{\cwjuT}{{\overset{\boldsymbol{\sim}}{\mathbf{w}}\traspd{j_1}}}

\newcommand{\muju}{\mu_{j_1}}

\newcommand{\cX}{\overset{\sim}{\mathbf{X}}}

\newcommand{\cXjT}{\overset{\sim}{\mathbf{X}}\traspd{j}}

\newcommand{\dX}{{\overset{\boldsymbol{.}}{\mathbf{X}}}}

\newcommand{\bdS}{{\overset{\boldsymbol{.}}{\mathbf{S}}}}

\newcommand{\dSj}{{\overset{\boldsymbol{.}}{\mathbf{S}}_j}}
\newcommand{\dSjinv}{{\overset{\boldsymbol{.}}{\mathbf{S}}\matinvj{j}}}

\newcommand{\dXj}{{\overset{\boldsymbol{.}}{\mathbf{X}}_j}}
\newcommand{\dXjT}{{\overset{\boldsymbol{.}}{\mathbf{X}}\traspd{j}}}

\newcommand{\da}{{\overset{\boldsymbol{.}}{\mathbf{a}}}}
\newcommand{\daj}{{\overset{\boldsymbol{.}}{\mathbf{a}}_j}}
\newcommand{\dajT}{{\overset{\boldsymbol{.}}{\mathbf{a}}\traspd{j}}}

\newcommand{\dbj}{{\overset{\boldsymbol{.}}{\mathbf{b}}_j}}

\newcommand{\ddj}{{\overset{\boldsymbol{.}}{\mathbf{d}}_j}}
\newcommand{\ddjT}{{\overset{\boldsymbol{.}}{\mathbf{d}}\traspd{j}}}

\newcommand{\bQTj}[1]{{\mathbf{Q}}_{\mathbf{#1}_{[j-1]}}}
\newcommand{\bQTjT}[1]{{\mathbf{Q}}\traspsMb{#1}}
\newcommand{\dQTj}[1] {{\overset{\boldsymbol{.}}{{\mathbf{Q}}}_{\mathbf{#1}_{[j-1]} }}}

\newcommand{\hQTj}[1]{{\widehat{\mathbf{Q}}}_{{#1}_{[j-1]}}}
\newcommand{\hQTjT}[1]{{\widehat{\mathbf{Q}}}\traspsM{#1}}

\newcommand{\Qcj}{\mathbf{\cal{Q}}_j}
\newcommand{\QcjT}{\mathbf{\cal{Q}}\trasps{j}}
\newcommand{\dQcjT} {{\overset{\boldsymbol{.}}{\mathbf{\cal{Q}}}\traspd{j} }}
\newcommand{\dQcj}{{\overset{\boldsymbol{.}}{\mathbf{\cal Q}}_j}}

\newcommand{\vexp}{\mathrm{vexp}}
\newcommand{\vexpq}{\mathrm{vexp}_Q}
\newcommand{\evexp}{\mathrm{evexp}}
\newcommand{\cvexp}{\mathrm{cvexp}}
\newcommand{\rCvexp}{\mathrm{rCvexp}}

\newcommand{\trace}{\mathrm{tr}}
\newcommand{\diag}{\mathrm{diag}}

\DeclareMathOperator*{\argmin}{arg\,min}
\DeclareMathOperator*{\argmax}{arg\,max}

\newcolumntype{C}{@{\extracolsep{0.0cm}}l@{\extracolsep{0.25cm}}}%

\newcommand{\blind}{0}

\begin{document}
%
\begin{frontmatter}
\title{Projection sparse principal component analysis: \\
An efficient least squares method}
\if0\blind{
\author[A1]{Giovanni Maria Merola\corref{cor1}}
\author[A1,A2]{Gemai Chen}
\address[A1]{Department of Mathematical Sciences, Xi'an Jiaotong-Liverpool University,
111 Ren€™ai Road,\\Suzhou Industrial Park, Suzhou, Jiangsu Province, P.R. China 215123}
\address[A2]{Department of Mathematics and Statistics, University of Calgary, Calgary, Alberta, Canada T2N IN4 }
\cortext[cor1]{Corresponding author. Email address: \url{giovanni.merola@xjtlu.edu.cn}}
}\fi

\begin{abstract}
We propose a new sparse principal component analysis (SPCA) method in which the solutions are obtained by projecting the full cardinality principal components onto subsets of variables. The resulting components are guaranteed to explain a given proportion of variance. The computation of these solutions is very efficient. The proposed method compares well with the optimal least squares sparse components.
We show that other SPCA methods fail to identify the best sparse approximations of the principal components and explain less variance than our solutions. We illustrate and compare our method with others with extensive simulations and with the analysis of the computational results for nine datasets of increasing dimensions up to 16,000 variables.
\end{abstract}

\end{frontmatter}

\section{Introduction}

Principal components analysis (PCA) is the oldest and most popular data dimensionality reduction method used to approximate a set of variables in a lower dimensional space \cite{pea}. Effective use of the method can approximate a large number of variables by a few linear combinations of them, called principal components (PCs). PCA has been extensively used over the past century and in recent times the interest in this method has surged, due to the availability of very large datasets. Applications of PCA include the analysis of gene expression analysis, market segmentation, handwriting classification, image recognition, and other types of data.

PCs are usually difficult to interpret and not informative on important features of the dataset because they are combinations of all the observed variables, as already pointed out by \citet{jef}. A common approach used to increase their interpretability is to threshold the coefficients of the combinations defining the PCs, which are called loadings. That is, variables corresponding to loadings that are lower than a given threshold are ignored. However, this practice can give misleading results \citep{jol} and the retained variables can be highly correlated among themselves. This means that the variables included in the interpretation actually carry similar information.

In recent years a large number of methods for sparse principal components analysis (SPCA) have been proposed; see, e.g., \citep{jol00, mog, zou, sri, she, wan}. These methods compute solutions in which some of the coefficients to be estimated are equal to zero. In addition to increased interpretability of the results, sparse methods are recommended under the sparsity principle \citep{has}.

Conventional SPCA methods replace the ordinary PCs with PCs of subsets (blocks) of variables. The resulting sparse PCs (SPCs) are combinations of only a few of the observed variables. That is, the SPCs are linear combinations of all the variables with only few loadings not equal to zero, the number of which is called cardinality. The difference among conventional SPCA methods is in the optimization approach used to select the variables to be included in the blocks. In this context, the variable selection problem is non-convex NP-hard \citep{mog}, hence computationally intractable. Some methods use a genuine cardinality penalty (improperly called $\ell_0$ norm), others an $\ell_1$ penalty. The most popular of these methods seems to be the Lasso based SPCA \citep{zou}.

SPCA methods are expressly recommended for large fat datasets \citep{has2}, i.e., samples with fewer observations than variables. By the nature of the objective function maximized, the components computed maximize the variance explained of each block, instead of that of the whole data matrix. As a consequence, the selected blocks contain highly correlated variables \citep{mer}. Furthermore, with this approach the exact sparse reduction of the PCs of fat matrices cannot be identified, as we will show later.

A least squares SPCA method (LS SPCA) in which the sparse components are obtained by minimizing the $\ell_2$ norm of the approximation error was  proposed in \cite{mer}.
This approach produces sparse components that explain the largest possible proportion of variance for a given cardinality. LS SPCA can identify the equivalent sparse representation of the PCs of fat matrices. However, the variable selection approaches suggested are not scalable to large matrices because they are top-down and require the computation of (generalized) eigenvectors of large matrices.

In this paper we suggest an efficient variable selection strategy for LS SPCA, based on projecting the full cardinality PCs on blocks of variables. This approach is based on a property we prove which says that if the regression of a PC on a block of variables yields an $R^2$ statistics equal to $\alpha \in (0, 1)$, then the LS SPCA components computed on that block of variables will explain a proportion of the variance not smaller than $\alpha$.
With this approach, the NP-hard SPCA variable selection problem is reduced to a more manageable univariate regression variable selection problem. This procedure, to which we refer to as Projection SPCA (PSPCA), also gives as by-products the projections of the PCs, which are sparse components in themselves.

We show that algorithms using PSPCA variable selection are very efficient for computing LS SPCA components, having a growth order of about the number of variables squared. We also show that the  performance of the projected PCs is comparable to that of the LS SPCA components. This is relevant because these projections are easier to understand by researchers and also easier and more economical to compute.

In the next section we review PCA and LS SPCA, and give a novel interpretation of the latter. The methodological details of PSPCA are discussed in Section~\ref{sec:pspca}. In Section~\ref{sec:selectblocks} we discuss the use of PSPCA for variable selection. We compare its performance on fat matrices with that of conventional SPCA methods and explain the details of the PSPCA algorithm. The proposed method is illustrated by using simulated and real datasets in Section~\ref{sec:examples}. We give some final comments in Section~\ref{sec:conclusions}. The Appendix contains some of the proofs.

\section{Full cardinality and sparse principal components}\label{sec:sparsecomp}

We assume that $\bX$ is an $n\times p$ matrix containing $n$ observations on $p$ mean centred variables which have been scaled so that $\bS = \bX\trasp\bX$ is the sample covariance or correlation  matrix. Because of this, we will use the terms uncorrelated (correlated) and nonorthogonal (orthogonal) interchangeably. The information contained in the dataset is summarized by its total variance, defined by the squared (Frobenius) norm $||\bX||^2 = \trace(\bS)$, where $\trace$ is the trace operator. In the following the term norm refers to this norm, unless otherwise specified.

A  component is any linear combination of the columns of $\bX$, generically denoted by $\bt = \bX\ba$, where the vector $\ba$ is the vector of loadings (or just the loadings, for short). A set of ordered components $(\bt_1, \ldots, \bt_d) = \bX(\ba_1, \ldots, \ba_d)$ is denoted as $\bTd = \bX\bAd$, where the subscript ${[j]}$ denotes the first $j$ columns of a matrix and $\bAd = (\ba_1, \ldots, \ba_d)$ is called the matrix of loadings.

The least squares estimates of $d$ components, with $d \le p$, are obtained by minimizing the squared norm of the difference of the data matrix from its projection onto the components, $\Pi_{\bTd}\bX$, where $\Pi_{\bM}=\bM(\bM\trasp\bM)\matinv\bM\trasp$ denotes the projector onto the column space of the matrix $\bM$. By Pythagoras' theorem, the solutions must satisfy
\begin{equation}\label{eq:vexp}
\bTd = \argmin\limits_{\bM\in \Re^{n\times d}}   ||\bX - \Pi_{\bM}\bX||^2 = \argmax\limits_{\bM\in \Re^{n\times d}} ||\Pi_{\bM}\bX||^2.
\end{equation}
The term on the right-hand side of Eq.~(\ref{eq:vexp}), called the variance explained by the components $\bTd$ and denoted as $\vexp(\bTd)$, is used to measure the performance of the components in approximating the data and is equal to
\begin{align}
\vexp(\bTd) = \trace \Big \{ \bX\trasp\bTd (\bTdT\bTd )^{-1}\bTdT\bX \Big \} = \trace \Big \{ \bS\bAd (\bAdT\bS\bAd )^{-1}\bAdT\bS \Big \}.
\label{eq:vexp1}
\end{align}

\subsection{Principal components}

The principal components, denoted as $\bu_j = \bX\bv_j$ with $j \in \{ 1,\ldots, d\}$ and $d \leq \mathrm{rank}(S)$, are obtained by maximizing $\vexp(\bTd)$ under the orthogonality requirements $\bu\trasps{i}\bu_j = 0$ if $i\neq j$. That is, the PCs' loadings are found from
\begin{align}\label{eq:pcaprob}
\forall_{j \in \{ 1, \ldots, d\}} \quad  \bvj = &\argmax\limits_{\baj\in\Re^p} {\bajT\bX\trasp\bX\bX\trasp\bX\baj}/{\bajT\bXT\bX\baj}, \\
  &\text{subject to}\, \bv\trasps{i}\bS\baj=\mathbf{0},\, i < j, \,\text{if } j >1.\notag
\end{align}
The solution loadings $\bv_j$ are the eigenvectors of $\bS$, such that $\bS\bv_j = \bv_j\lambda_j$, corresponding to the eigenvalues in nondecreasing order, $\lambda_1 \geq  \cdots \geq \lambda_d$. By the orthogonality of the components, the total variance explained  can be broken down as the sum of the individual variances explained as
$$
\vexp\left(\bU_{[d]}\right) = \sum_{j=1}^d\vexp(\buj) = \sum_{j=1}^d {\bujT\bX\bXT\buj}/{\bujT\buj}
= \sum_{j=1}^d \lambda_j.
$$

The PCs can be regarded as solutions to a number of different problems, such as the singular value decomposition \citep{eck}. Most notably, \citet{hot} showed that when the loadings are scaled to unit norm, we have $\vexp(\bu_j) =\bu\trasps{j}\bu_j = \lambda_j$. Noting also that, by the properties of the spectral decomposition of a symmetric matrix, the loadings are orthogonal, then the PCA problem can be formulated as finding
\begin{align}\label{eq:pcahot}
 \forall_{j \in \{ 1, \ldots, d\}} \quad   \bv_j =& \argmax_{\ba_j\in\Re^p} \ba\trasps{j}\bS\ba_j, \\
    &\text{subject to } \ba\trasps{j}\ba_j = 1, \ba\trasps{j}\bv_i = 0,\, \text{if}\ j > i \geq 1.\nonumber
\end{align}

\subsection{Least squares sparse principal components}

A sparse component is a linear combination of a subset, $\dX_j$, of columns of the $\bX$ matrix, or block of variables, defined by the sparse loadings $\da_j$ as $\bt_j = \dX_j\da_j$. The number of variables in the block is the cardinality of the loadings. The least squares sparse PCA (LS SPCA) problem is defined by adding sparsity constraints directly into the PCA objective \eqref{eq:pcaprob}, which gives
\begin{align}\label{eq:spcaprob}
 \forall_{j \in \{ 1, \ldots, d\}} \quad  \dbj =& \argmax\limits_{\daj\in\Re^{c_j}} {\dajT\dXjT\bX\bX\trasp\dXj\daj}/{\dajT\dXjT\dXj\daj} \\
  & \text{subject to}\, \bb\traspd{i}\bS\baj = 0,\, i < j, \,\text{if } j >1,\nonumber
\end{align}
where $\dXj$ is a block of $c_j$ variables, $\ba_j = \bJ_j\daj$ and $\bbj = \bJ_j\dbj$ are the full cardinality representations of the sparse loadings, defined by means of the matrix $\bJ_j$, which is formed by the columns of the order-$p$ identity matrix corresponding to the variables in $\dXj$. Note that the SPCA objective must be maximized sequentially. We will refer to the components in the order with which they are computed. The solutions are given in the following proposition \citep{mer}.

\begin{proposition}[Uncorrelated LS SPCA]\label{prop:uspca}
Given a block of $c_j$ linearly independent variables $\dX_j$, the solutions to objective \eqref{eq:spcaprob} are the generalized eigenvectors satisfying
\begin{equation}\label{eq:uspca}
\bC_j\dXjT\bX\bXT\dXj\dbj = \bC_j\bJ\trasps{j}\bS\bS\bJ_j\dbj = \dSj\dbj\xi_{j},
\end{equation}
where
$$
\bC_j = \{ \bI_{c_j} -  \bHjT(\bHj\bdS\matinvj{j}\bHjT)\matinv\bHj\bdS\matinvj{j} \} , \quad \bC_1 = \bI_{c_1},
$$
the sparse loadings $\dbj$, the orthogonality constraints $\bHj =  \bY\trasps{[j-1]}\dXj$, with $\bdS_j = \dXjT\dXj$ and $\xi_{j} = \vexp(\by_j)$ the largest eigenvalue. The SPCs $\by_j = \dXj\dbj$ are mutually orthogonal and maximize the variance explained.
\end{proposition}

The optimal orthogonal LS SPCA components are highly constrained; for example their cardinality cannot be smaller than their order. Due to the greedy nature of the optimization carried out, these locally optimal orthogonal components often stride away from the global optimum, while globally better solutions can be found by removing the orthogonality constraints \cite{mer}.

If the orthogonality constraints are dropped, the net increment in total variance explained due to a new component is the variance explained by the residuals orthogonal to the components already in the model. This extra variance explained, which we denote as $\evexp$, is equal to
\begin{align}
    \evexp(\btj) &= ||\Pi_{[\bQTj{T}\baj]}\bX||^2 =
    {\bajT\bQTjT{T}\bX\bX\trasp\bQTj{T}\baj}/{\bajT\bQTjT{T}\bQTj{T}\baj}\nonumber\\
    &=  {\bajT\bX\trasp\bQTj{T}\bQTjT{T}\bX\baj}/{\bajT\bQTjT{T}\bQTj{T}\baj},
\label{eq:evexp}
\end{align}
where $\bQTj{T} = (\bI_n - \Pi_{\bT_{[j-1]}})\bX$ is the orthogonal complement of the $\bX$ matrix ($\bQ_{T_{[0]}} = \bX$), to which we refer as deflated $\bX$ matrix (with respect to $\bT_{[j-1]}$), and $\bQTj{T}\ba_j = \bt_j - \Pi_{\bT_{[j-1]}} \bt_j$ is the orthogonal residual of $\bt_j$.
For the first component $\evexp(\bt_1) = \vexp(\bt_1)$. The total variance explained by a set of correlated components is equal to the sum of the extra variances explained, viz.
$$
\vexp(\bTd) = \sum_{j = 1}^d \evexp(\bt_j).
$$

The sparse solutions cannot be determined from the maximization of objective \eqref{eq:evexp} because this is defined in terms of the deflated components $\bQTj{T}\ba_j$, while the cardinality constraints must be imposed on the $x$-variables.

For this reason, \citet{mer} derives nonorthogonal SPCs $\bz_j = \bX\bd_j = \dXj\ddj$ which maximize the variance of $\bQTj{Z}$ explained by a component $\bz_j$. This is defined in terms of the variables $\dXj$ as
\begin{equation}\label{eq:vexpq}
\vexp_Q(\bzj) =  ||\Pi_{\bz_j}\bQTj{Z}||^2 = {\bz\trasps{j}\bQTj{Z}\bQTjT{Z}\bzj}/{\bz\trasps{j}\bzj}
= {\ddjT\dXjT\bQTj{Z}\bQTjT{Z}\dXj\ddj}/{\ddjT\dXjT\dXj\ddj}.
\end{equation}

\begin{proposition}[Correlated LS SPCA]
Given a block of linearly independent variables $\dX_j$, the correlated SPCs, $\bz_j = \dXj\ddj = \bX\bd_j$, that successively maximize $\vexpq$ are the generalized eigenvectors satisfying
\begin{equation}\label{eq:lsspcasol}
    \dXjT\bQTj{Z}\bQTjT{Z}\dXj\ddj = \dXjT\dXj\ddj\gamma_j = \dSj\ddj\gamma_j,
\end{equation}
where  $\gamma_j$ is the largest generalized eigenvalue, which is equal to $\vexpq(\bzj)$. The full cardinality loadings are equal to $\bd_j = \bJ_j\ddj$. The first component is identical to the first orthogonal component.
\end{proposition}

We will refer to the SPCs derived from the minimization of the least squares criterion generically as LS SPCA components and use USPCA and CSPCA to refer to the uncorrelated and correlated solutions, respectively.

The variance of $\bQ_{\bT_{[j]}}$ that a component explains is a lower bound for the variance of $\bX$ that this component can explain, as stated in the following proposition, whose proof is given in the Appendix.

\begin{proposition}\label{prop:vqltev}
Given an ordered set of $d$ components, $\bt_j = \bX\ba_j, j = 1, \ldots, d$,
the different types of variance explained as defined in Eqs.~\eqref{eq:vexp1}, \eqref{eq:evexp} and \eqref{eq:vexpq} satisfy
\begin{equation}\label{eq:ineqvexp}
\vexp_Q(\btj) \leq \evexp(\btj) \leq \vexp(\btj),
\end{equation}
where $\vexp(\btj) = {\btjT\bX\bXT\btj}/{\btjT\btj}$.
Equality is achieved for the first component or if a component is orthogonal to the preceding ones.
\end{proposition}

The difference between $\evexp$ and $\vexpq$ lies in the different spaces onto which the matrix $\bQTj{T}$ is projected. The extra variance explained measures the norm of the projection of $\bQTj{T}$ onto a component in the span of
$$
\cspace(\dQTj{T})\subseteq \cspace(\bQTj{T}).
$$
Instead, $\vexpq$ measures the norm of the projection onto a component in the span of
$$
\cspace(\dXj) = \cspace(\dQTj{T} + \Pi_{T_{[j-1]}}\dXj)\nsubseteq  \cspace(\dQTj{T}),
$$
where $\cspace(\bA)$ denotes the column space of the matrix $\bA$. This leads to the simple interpretation of the LS SPCA solutions as the first PCs of two different projections of the $\bX$ matrix, as shown in the next theorem, the proof of which is in the Appendix.

\begin{theorem}\label{th:spcaisproj}
Let $\dXj$ be a block of linearly independent variables. Then,
\begin{enumerate}
\item[(i)] The orthogonal LS SPCA components, $\by_j = \dXj\dbj$, are the first PCs of the matrices $ ( \Pi_{\dXj} - \Pi_{\hat{Y}_\dXj} ) \bX$, where
$$
\hat{Y}_\dXj = \Pi_{\dXj}\bY_{[j-1]}.
$$
\item[(ii)]
The nonorthogonal LS SPCA components, $\bzj = \dXj\ddj$, are the first PCs of the matrices
$$
\hQTj{Z} = \Pi_{\dXj}\bQTj{Z} = \Pi_{\dXj} (\bI - \Pi_{\bZ_{[j-1]}} )\bX.
$$
\end{enumerate}
\end{theorem}

\subsection{Conventional sparse principal components}

Conventional SPCA methods compute the sparse components as the first PCs of blocks of variables deflated in different ways~\citep{mog}. These solutions are derived with different motivations, either from a constrained LS approach (see, among others, \citep{zou, she}) or by directly adding sparsifying penalties to the Hotelling's formulation of PCA in Eq.~(\ref{eq:pcahot}); see \citep{mer} for a discussion. Hence, the SPCs are the PCs of the (possibly deflated) blocks of variables and the loadings are the solution to
\begin{align*}
\max\limits_{\baj} \bajT\QcjT\Qcj\baj \quad \Leftrightarrow \quad \max\limits_{\daj} \dajT\dQcjT\dQcj\daj, \quad \mathrm{card}(\baj) = c_j,
\quad \bajT\baj=1  \quad   \dajT\daj=1,
\end{align*}
where $\Qcj$ denotes the $\bX$ matrix deflated using one of the different existing methods; for a review, see \citep{mac}. In conventional SPCA the norm of the components is considered equivalent to the variance of $\bX$ that they explain. It is easy to show that this latter assumption is not true \citep{mer}, thus these components do not maximize the variance explained. Furthermore, the blocks selected will contain highly correlated variables because the more correlated the variables in the block are, the larger the first eigenvalue of their covariance (or correlation) matrix. Hence, conventional SPCs will have larger cardinality and explain less variance than LS SPCs.

\section{Projection sparse principal components}\label{sec:pspca}

The idea underpinning PSPCA is to iteratively project the (full cardinality) first principal components of the deflated matrices onto blocks of variables $\dXj$. These projections, to which we refer to as projection sparse components, are SPCs in themselves and the variance of the PCs that they explain is a lower bound for the extra variance of $\bX$ explained by an LS SPCA component, as we prove next.

Let $\bQTj{\bhR}$ denote the $\bX$ matrix deflated  of the first $j-1$ projection SPCs, $\hr_{i_1}$ for all $i  \in \{ 1,\ldots, j-1\}$, and
$$
\bQTjT{\bhR}\bQTj{\bhR} = \bW_j\bMj\bW\trasps{j},
$$
with $\bMj = \diag (\mu_{j_1} \geq\cdots\geq \mu_{j_p})$ the eigendecomposition of its covariance matrix, the PCs of $\bQTj{\bhR}$ are
$$
\bR_j = \bQTj{\bhR}\bW_j.
$$

Since the PCs, $\br_{j_i}$, are orthogonal to the previously computed components, then $\vexp(\brju) = \evexp(\brju) = \brjuT\brju= \muju$. Hence, $\vexp(\brju)$ is an upper bound for the extra variance explained by any component, $\bt_j = \bX\ba_j$, added to the model, i.e., $\evexp(\bt_j)\leq \muju$.

Assume that the variables in a block $\dXj$ are linearly independent and explain a proportion of the variance of $\brju$ not less than $\alpha \in (0,1)$, i.e., $\hrju = \Pi_{\dXj}\brju$ is such that
\begin{equation}\label{eq:rsqu}
{\hrjT\hrju}/{\brjuT\brju} \geq \alpha\; \mathrm{or, equivalently,} \; \hrjuT\hrju \geq \alpha\mu_{j}.
\end{equation}
Since $\cspace(\bQTj{\bhR})\subseteq \cspace(\bX)$, a subset of variables $\dXj$ satisfying (\ref{eq:rsqu}) can be found for any $\alpha\in[0,1]$. The projection $\hrju$ is an SPC defined by $\hrju = \dXj\hwju$, with loadings
\begin{equation}\label{eq:pspcaloads}
    \hwju = (\dXjT\dXj)^{-1}\dXjT\brju.
\end{equation}
A lower bound for the extra variance explained by $\hrju$ is given in the following theorem.

\begin{theorem}\label{th:prspca}
Let $\hrju$ be the projection of the first PC of $\bQTj{\bhR}$ on a block of variables $\dXj$ such that $\hrjuT\hrju\geq \alpha\muju$. Then, $\evexp(\hrju) \geq \alpha\muju$.
\end{theorem}

\medskip
\noindent
\textbf{Proof}.
By substituting the eigendecomposition $\bQTj{\bhR}\bQTjT{\bhR} = \bRj\bR\trasps{j}$ into Eq.~(\ref{eq:vexpq}), it is easy to verify that
\begin{equation*}\label{eq:evexprhat}
\evexp(\hrju) \geq \vexpq(\hrju) =
\hrjuT\hrju + \sum_{i>1}  {(\hrjuT\brji)^2}/{\hrjuT\hrju} \geq
\hrjuT\hrju \geq \alpha \mu_{j_1},
\end{equation*}
because of Eqs.~\eqref{eq:ineqvexp} and \eqref{eq:rsqu}.\hfill $\Box$

\bigskip
The intricacy of the iterated projections renders the comparison between SPCs computed with different methods difficult. In the  following theorem we show that $\hrjuT\hrju$ is a lower bound for the extra variances explained by the LS SPCs.

\begin{theorem}\label{th:ineqvexp}
Let $\dXj$ be a block of linearly independent variables and $\hrjuT\hrju\geq \alpha\muju$. Also let $\bz_j$ and $\by_j$ be the correlated and uncorrelated SPCA components, respectively, and assume that, when $j>1$, all components have been computed with respect to the same set of previous components $\bT_{[j-1]}$. Then, the following properties hold:
    \begin{subequations}\label{eq:th3}
        \begin{equation}\label{th3:i}
            \alpha \mu_{j}\leq  \vexpq(\hrju) \leq \evexp(\hrju)\leq \evexp(\byj)\leq \mu_j,
        \end{equation}
         \begin{equation}\label{th3:ii}
          \alpha \mu_{j}\leq  \vexpq(\hrju) \leq  \vexpq(\bzj) \leq \evexp(\bzj)\leq \evexp(\byj)\leq \mu_j.
        \end{equation}
    \end{subequations}
\end{theorem}

\medskip
\noindent
\textbf{Proof}.
By definition, $\byj$ is the linear combination of the variables in $\dXj$ that explains the most possible extra variance of $\bX$ and inequality (\ref{th3:i}) follows from Theorem~\ref{th:prspca}. By substituting the PCA decomposition into Eq.~\eqref{eq:vexpq}, it can be verified that
\begin{equation}\label{eq:vlspr}
\vexpq(\bzj) = \max\limits_{\bt_j=\dXj\daj}\sum_{i = 1}^p {(\bt\trasps{j}\br_{j_i})^2}/{\bt\trasps{j}\bt_j}
\geq \sum_{i = 1}^p {(\hrjuT\br_{j_i})^2}/{\hrjuT\hrju} =  \vexpq(\hrju) \geq \alpha\mu_j.
\end{equation}
This, together with the optimality of $\evexp(\byj)$,  proves inequality (\ref{th3:ii}).
When considering the first components, the inequalities reduce to $\alpha \mu_{1}\leq  \evexp(\hr_{1_1})  \leq \evexp(\bz_1)= \evexp(\by_1)\leq \mu_1$, because for the first components $\vexp(\bt_1) = \vexpq(\bt_1) = \evexp(\bt_1)$.\hfill $\Box$

\bigskip
In principle it cannot be excluded that $\evexp(\hrju) > \evexp(\bzj)$.
The question of how different are $\hrju$ and $\bzj$ when computed with respect to the same deflated matrix $\bQj$ does not have a straightforward answer. We have that $\bzj$ is the linear combination of the variables in $\dXj$ which maximizes $\vexpq$, while $\hrju$ is the component most correlated with the first PC $\brju$. Given that $\bt_j= \dXj\da_j$, $ {(\bt\trasps{j}\br_{j_i})^2}/{\bt\trasps{j}\bt_j} = \mathrm{corr}^2(\bt_j, \br_{j_i})\mu_{j_i}$,
from Eq.~(\ref{eq:vlspr}) it follows that
\begin{align*}
  \vexpq(\bzj)-\vexpq(\hrju) &= \sum_{i=1}^p (\beta_{j_i} - \alpha_{j_i})\mu_{j_i}
  = \sum_{i>1} (\beta_{j_i} - \alpha_{j_i})\mu_{j_i} -  (\alpha_{j_1} - \beta_{j_1})\mu_{j_1}\geq 0,
\end{align*}
where $\beta_{j_i} = \mathrm{corr}^2(\bz_j, \br_{j_i})$ and  $\alpha_{j_i} = \mathrm{corr}^2(\hrju, \br_{j_i})$, and necessarily
$\alpha_{j_1} \geq \beta_{j_1}$.

The following lemma, which is proved in the Appendix, is useful to characterize the difference in variance of $\bQ_j$ explained by the two components.

\setcounter{lemma}{0}
\begin{lemma}\label{le:corr}
Let $t$ and $x$ be two random variables and $\by = ( y_1, \ldots, y_d)\trasp$ a set of $d$ random variables uncorrelated with $x$. If $\mathrm{corr}^2(t, x) =\alpha$, then for all $i \in \{ 1, \ldots, d\}$, $\mathrm{corr}^2(t, y_i) \leq 1 - \alpha$. If the $y_i$ variables are mutually uncorrelated, it follows that
    \begin{equation*}\label{eq:lemma1ii}
    \sum_{i=1}^d \mathrm{corr}^2(t, y_i) \leq 1 - \alpha.
     \end{equation*}
\end{lemma}

Since the PCs $\brji$ are mutually uncorrelated, assuming that $\beta_{j_1} > \beta_{j_2}$, by the inequalities in Lemma \ref{le:corr},
$$
0\leq \max \{\vexpq(\bzj) - \vexpq(\hrju)\} \leq \beta_{j_1}\mu_\ju + (1 - \beta_\ju)\mu_\jind{2} - \alpha_\ju\mu_\ju = (1-\beta_\ju)\mu_\jind{2} -(\alpha_\ju - \beta_\ju)\mu_\ju.
$$
Therefore, the squared correlations $\alpha_\ju$ and $\beta_\ju$ are such that
$$
\alpha_\ju - {\mu_\jind{2}(1 - \alpha_\ju)}/{(\mu_\ju - \mu_\jind{2})}
\leq \beta_\ju \leq  \alpha_\ju.
$$
Hence, when $\alpha_\ju$ is large and the eigenvalues $\mu_\ju$ and $\mu_\jind{2}$ are well separated, $\hrju$ and $\bzj$ will be very close because they have similar correlation with $\brju$. In our studies we found that the extra variance explained by the PSPCA components and the LS SPCA components is very similar. This is to be expected because the PSPCA components have the largest possible correlation with the first PCs $\brju$, which are the components that explain the most extra variance.

It should be noted that the inequalities in Theorem~\ref{th:ineqvexp} apply when the different components are computed after the same set of previous components. This is hardly possible in practice because the solutions computed with the various methods are (slightly) different, and different optimality paths are determined by greedy algorithms.

\section{Using projection SPCA to select the variables for the sparse components}\label{sec:selectblocks}

Finding an efficient algorithm for the selection of the variables forming the SPCs is fundamental for the scalability of an SPCA algorithm. Greedy approaches are required because searching the $2^{(p-1)d}$ possible subsets of indices for $d$ SPCs of unknown cardinality is a non-convex NP-hard, hence computationally intractable, problem. A first simplification adopted by most SPCA methods is to select the blocks of variables sequentially for each component. Also in this case, each problem is NP-hard \citep{mog}. The several greedy solutions proposed for conventional SPCA cannot be used for the LS SPCA problem because they seek subsets of variables that are highly correlated. \citet{mer} suggested a branch-and-bound and a backward selection algorithms for LS SPCA. Neither of these is efficient because they are top-down and require the computation of SPCs of large cardinality to evaluate the variance explained.

PSPCA provides a simple yet effective supervisor for the selection of subsets of variables for the computation of LS SPCs. In fact, by Theorem~\ref{th:ineqvexp}, it is enough to select a block of variables that explains a given percentage of the current first PC to be guaranteed that an LS SPCA component computed on that block will explain more than that percentage of the variance of the whole data matrix. Hence, by using PSPCA the LS SPCA variable selection problem is transformed into a more economical regression variable selection problem, thus eliminating the need of computing costly SPCs to evaluate the objective function (the variance explained by the SPCs).

Regression model selection has been extensively researched and several approaches for this task have been proposed. Any of these approaches can be used to select the blocks of variables, including Lasso and regularized lasso, if preferred. The regression approach has also the advantage of being familiar to most data analysts and  provides the projection  SPCs as a by-product.

\subsection{Comparison with conventional SPCA methods on ``fat'' matrices}\label{sec:convspca}

A particular concern  with conventional SPCA methods is that their objective function increases even when perfectly correlated variables are added to the model. Therefore, another important advantage of using a regression model selection method is that the blocks of variables can be chosen to have full column rank and not contain highly correlated variables. This property is important because parsimonious approximations of the PCs should not contain redundant variables. Hence, this property is a further reason for preferring the LS SPCA to conventional SPCA methods which, conversely, generate solution from blocks of highly correlated variables.

Another drawback of conventional SPCA methods, connected with the concern mentioned above, is that they cannot identify the most sparse representation of the PCs when applied to column rank deficient matrices. ``Fat'' matrices, i.e., datasets made up of more features than objects, are very common in the analysis of gene expression microarrays or near-infrared spectroscopy data, for example. In this case, the features are linearly dependent and the PCs can be expressed as linear combinations of as many variables as the rank of the matrix, as stated in the following lemma; see the Appendix for a proof.

\begin{lemma}
When $\mathrm{rank}(\bX) = r < p$ the principal components can be expressed as sparse components of cardinality $r$ and loadings that have norm larger than $1$.
\end{lemma}

When applied to column rank deficient matrices, conventional SPCA methods compute components of cardinality larger than the rank of the matrix because the only components with unit norm loadings that are equal to the PCs are the full cardinality PCs themselves. This fact is well documented by several examples available in the SPCA literature; see, e.g., \citep {wan,zou}. This means that the model is overfitted by the inclusion of redundant perfectly correlated variables.

When $\mathrm{rank}(\bX) = r$, the LS SPCA components computed on a block of $r$ independent variables will be equal to the full cardinality PC, because of Theorem~\ref{th:spcaisproj}. The same is true for PSPCA, when $r$ variables are enough to explain 100\% of the variance of the PCs. In fact, LS SPCA and PSPCA components of cardinality larger than the rank of the data matrix cannot be computed because in that case a matrix $\dXjT\dXj$ would be singular.

The following example shows the overfitting resulting from applying conventional SPCA to a matrix with perfectly correlated variables.
Consider a matrix with 100 observations on five perfectly collinear variables defined, for all $i \in \{ 1, \ldots, 100\}$ and $j \in \{ 1, \ldots, d\}$, by
\begin{equation*}
x_{ij} = (-1)^{i}\sqrt{j}.
\end{equation*}
The covariance matrix of these variables, $\bS = \bX\trasp\bX$, has rank $1$, and the only nonzero eigenvalue is equal to 1500. The first PC explains all the variance and can be written in terms of any of the variables as $\bx_j\sqrt{1500/s_{jj}}$ with $j \in \{ 1, \ldots, 5\}$, i.e., as cardinality $1$ components with loading larger than $1$.

The loadings, norms and relative norms (norm of a component/total variance) of the conventional SPCA optimal solutions for the complete set of conventional SPCs of increasing cardinality are shown in Table~\ref{tab:expl}. We see that $x_5$ ``explains'' just 33\% of the norm of the PC, $x_4$ and $x_5$ together only 60\% (with a net increase equal to 27\%) and so on. These results suggest that the variables with larger variances explain more variance than other collinear variables and that only the full cardinality PC explains the maximum variance. This is true because the norm of a component with unit norm loadings is bounded by $\trace(\dXjT\dXj)$.

\begin{table}[t!]
 \centering
  \caption{Loadings and norms of the conventional SPCA solutions for the covariance matrix.}

  \bigskip
\begin{tabular}{lrrrrr}
      & \multicolumn{5}{c}{Cardinality} \\
\cmidrule{2-6}
Variable     & 1     & 2     & 3     & 4     & 5 \\
\midrule
$x_1$ & 0     & 0     & 0     & 0     & 0.58 \\
$x_2$ & 0     & 0     & 0     & 0.60 & 0.52 \\
$x_3$ & 0     & 0     & 0.65 & 0.53 & 0.45 \\
$x_4$ & 0     & 0.75 & 0.58 & 0.46 & 0.37 \\
$x_5$ & 1     & 0.67 & 0.50 & 0.38 & 0.26 \\
\midrule
Norm  & 500   & 900   & 1200  & 1400  & 1500 \\
Rel norm & 0.33  & 0.60  & 0.80  & 0.93  & 1.0 \\
\bottomrule
\end{tabular}
\label{tab:expl}
\end{table}

The results of applying conventional SPCA to the correlation matrix of the variables in the above example is even more revealing.
The correlation matrix is a matrix of $1$s with only one nonzero eigenvalue equal to 5. The conventional SPCA optimal solutions are shown in Table~\ref{tab:expl2}. The results are given irrespectively of which variables are included, because the standardized variables are identical. The cardinality five component is the first PC, which explains all the variability. These results lead to the absurd conclusion that a linear combination of identical variables explains more variance than just one of them.

\begin{table}[t!]
\centering
  \caption{Loadings and norms of the conventional SPCA solutions for the correlation matrix.}

\bigskip
\begin{tabular}{lrrrrr}
       & \multicolumn{5}{c}{Cardinality} \\
\cmidrule{2-6}
& 1     & 2     & 3     & 4     & 5 \\
\midrule
Norm  & 1     & 2     & 3     & 4     & 5 \\
Rel norm & 0.2  & 0.4  & 0.6  & 0.8  & 1.0 \\
\bottomrule
\end{tabular}
\label{tab:expl2}
\end{table}

Applying LS SPCA to this dataset would yield a single cardinality one component both for the covariance and the correlation matrices, as expected. This can be seen by considering that the variance explained by any variable $x_j$ is equal to
$$
\vexp(\bx_j) =  {\bx\trasps{j}\bX\bX\trasp\bx_j}/{\bx\trasps{j}\bx_j} =
 {\sum_{i=1}^5 s_{ij}^2}/{s_{jj}} =
\sum_{i=1}^5 s_{ii} = \trace(S),
$$
because $\mathrm{corr}(x_i, x_j)^2 = s_{ij}^2/(s_{ii}s_{jj}) = 1$. Which variable is chosen depends on the algorithm used. The same is true for PSPCA because any variable explains 100\% of the variance of the first PC. An analogous example for blocks of collinear variables can be derived using the artificial data example introduced in \cite{zou}, as illustrated in \cite{mer}, where other unconvincing aspects of conventional SPCA are discussed.

\subsection{Computational considerations}\label{sec:compdet}

The basic algorithm for computing LS SPCA or PSPCA SPCs is outlined in Algorithm~\ref{algo:fspca}. The algorithm is straightforward and simple to implement.

The computation of the PSPCA loadings (line~\ref{algo:projection}) can be simplified as follows. Let us assume, without loss of generality, that the first $c_j$ variables in $\bX$ form the block $\dXj$ and that the block $\cX_j$ contains the remaining variables, so that $\bX = (\dXj, \cX_j)$. We write, correspondingly, $\bw_{j_1} = (\dwjuT, \cwjuT)\trasp$, then we have that
$$
\bXT\brju =
\begin{pmatrix}
  \dXjT\brju\\
  \cXjT\brju
\end{pmatrix} =
\begin{pmatrix}
  \dwju\mu_{j_1}\\
  \cwju\mu_{j_1}
\end{pmatrix},
$$
because $\bXT\brju = \bQTjT{\bhR}\bQTj{\bhR}\bw_{j_1} = \bw_{j_1}\mu_{j_1}$.

Substituting this expression into Eq.~(\ref{eq:pspcaloads}), we can write the PSPCA loadings $\hwj$ as
$$
\hwju = (\dXjT\dXj)^{-1}\dXjT\brju = (\dXjT\dXj)^{-1}\dwju\mu_{j_1}.
$$

The CSPCA loadings in Eq.~(\ref{eq:uspca}) can be computed as the generalized eigenvectors satisfying
$$
\bC_j\bJ\trasps{j}\bS\bS\bJ_j  \bC\trasps{j}\dbj = \dSj\dbj\xi_{j},
$$
as shown as Statement 1 in the Appendix.
The algorithm requires careful implementation because in its simplest form it is highly computationally demanding. The most demanding operations are the computation of the PCs of the $\bQ_j$ matrices, the extraction of submatrices (which is a costly operation when the number of variables is large), the deflation of the $\bX$ matrix, the multiplication $\bQTj{Z}\bQTjT{Z}$ (for CSPCA)  and the computation of generalized eigenvalues, when the cardinality is large.

The algorithm can be sped up by computing the first eigenvector of the deflated covariance matrix $\bQ\trasps{j}\bQ_j$ and, if necessary the generalized eigenvectors for the sparse loadings, with the iterative power method. The simple power method algorithm is not necessarily very efficient, especially when the first two eigenvalues are not well separated, and other more efficient but complex algorithms could be used, e.g., Lanczos iterations \citep{bjo} or LOBPCG \citep{kny}.

\begin{algorithm}[H]\caption{Projection LS SPCA}%
    \begin{algorithmic}[1]
    \Procedure {plsspca}{$\bX,\, \alpha,\, computePSPCA,\, computeCSPCA,
    stopRuleCompute$}
    \State \textbf{initialize}
        \State {\hspace{1em}$\bQ_1 \gets \bX$}
        \State{\hspace{1em}$j \gets 0$}
        \State{\hspace{1em}stopCompute $\gets$ FALSE}
    \State \textbf{end initialize}
    \While{(stopCompute = FALSE)}\Comment{\textbf{start components computation}}
        \State{$j \gets j + 1$}
        \State{$\brju = \bQj\bwju :\, \bQ_j\bQ\trasps{j}\brju = \brju\mu_j$} \Comment{compute first
        PC of $\bQj$}\label{algo:pwm}
        \State{$ind_j \gets \{i_1,\ldots, i_{c_j}\}:\, ||\hrj||^2 \geq
        \alpha\mu_j$} \Comment{variable selection output}\label{algo:fwdselect}
        \State{$\dXj\leftarrow \bX[,\, ind_j]$}\Comment{$\dX_j$ are columns of $\bX$ in
        $ind_j$}
        \If{(computePSPCA)}\Comment{PSPCA}
            \State{$\daj \gets (\dX\traspd{j}\dX_j)\matinv\dwju$}\label{algo:projection}
            \Comment{$\dwju$ are the elements of $\bwju$ in $ind_j$}
         \ElsIf{(computeCSPCA)}\Comment{Correlated LS SPCA}
           \State{$\daj: \dXjT\bQ_j\bQjT\dXj\da_j = \dXjT\dXj\daj\gamma_j$}
        \Else\Comment{Uncorrelated LS SPCA}
           \State{$\daj: \bC_j\dXjT\bXj\bXjT\dXj\daj = \dXjT\dXj\daj\xi_j$}
        \EndIf
        \State{$ \btj \gets \dX_j\daj$} \Comment{$j$-th sparse component}
        \If{(stopRuleCompute = FALSE)}\quad\Comment{stop rule on total variance explained or
        number of components}
            \State{$ \bQ_{j+1} \gets \bQj - \frac{\btj\btjT}{\btjT\btj}\bQj$}
            \Comment{deflate $\bX$ of current component}\label{algo:deflX}
            \State{$\cvexp(j) \gets \trace(\bX\trasp\bX) - \trace(\bQ\trasps{j+1}\bQ_{j+1})$}
            \Comment{cumulative \rm{vexp}}\label{algo:cvexp}
        \Else
            \State stopCompute $\gets$ TRUE \Comment{\textbf{terminate components
            computation}}
        \EndIf
    \EndWhile
    \EndProcedure
    \end{algorithmic}\label{algo:fspca}
\end{algorithm}

In our implementation we used the simple version of the power method, which has complexity growth rate of about $O(p^2)$, while direct algorithms that compute the whole set of eigenvectors accurately are about $O(p^3)$. The power method is used in extremely high dimensional problems (for example, Google page ranking \citep{bry}) and in various algorithms for conventional SPCA, including \citep{jou, wan}.

The computational complexity of the algorithm depends also on which variable selection algorithm is used (line \ref{algo:fwdselect}). For our implementation we chose the fast greedy forward selection in which the variables that explain the most extra variance conditionally on the variables already in the model are selected until a given percentage of variance is explained. This method can be seen as a QR decomposition with supervised pivoting and can be implemented efficiently using updating formulas; see, e.g., Section~2.4.7 in~\citep {bjo}. Since the QR decomposition can be stopped after $c_j$ iterations, identifying a block will be an operation of order about $O(2c_jnp)$.

When applied to fat matrices, the solutions can be computed more economically by using a ``reverse svd'' approach (see Section~12.1.4 in~\citep{bish}), which means computing the eigenvectors of $\bX\trasp\bX$ starting from the $\bX\bX\trasp$ matrix.

The theoretical time complexity of the PLSSPCA algorithms cannot be computed exactly because the time taken to compute the eigenvectors with power iterations (if used) and to select the variables and extract submatrices depends on the implementation and the structure of the data. However, we expect the order of growth of the whole algorithm to be not higher than the complexity of computing the first PC of $\bQj$ as it does not contain operations of higher complexity. Therefore, the computation of each vector of loadings should be about $O(p^3)$ when direct eigendecomposition of $\bQ_j$ is used and roughly $O(p^2)$ when the power method is used. In the next section we will analyze the run time empirically.

\section{Numerical results}\label{sec:examples}

In this section we report the results of running repetitions of USPCA, CSPCA and PSPCA on simulated and real datasets, each with different number of variables, from 100 to over 16,000. To compute the components we applied forward selection requiring that each sparse component explained at least 95\% of the corresponding PC's variance. Since the number of factors considered is large and displaying the results would require very large tables, we present the results mostly graphically, highlighting the main features.

The computational times reported were measured using an implementation of the algorithm in \textsf{C++} embedded in \textsf{R} using the packages \texttt{Rcpp} and  \texttt{RcppEigen}. The execution times were measured on an eptacore Intel$^\circledR$\ Core(TM) i7-4770S CPU @ 3.10GHz using Windows 7 operating system.  It is well known that \textsf{R} is an inefficient language \citep{mor}, so the run times are slower than what they would be if the programs had been written in a lower level language.

\subsection{Simulations}

\begin{figure}[b!]
\centering
  \includegraphics[width = 1\textwidth, height = 0.3\textwidth]{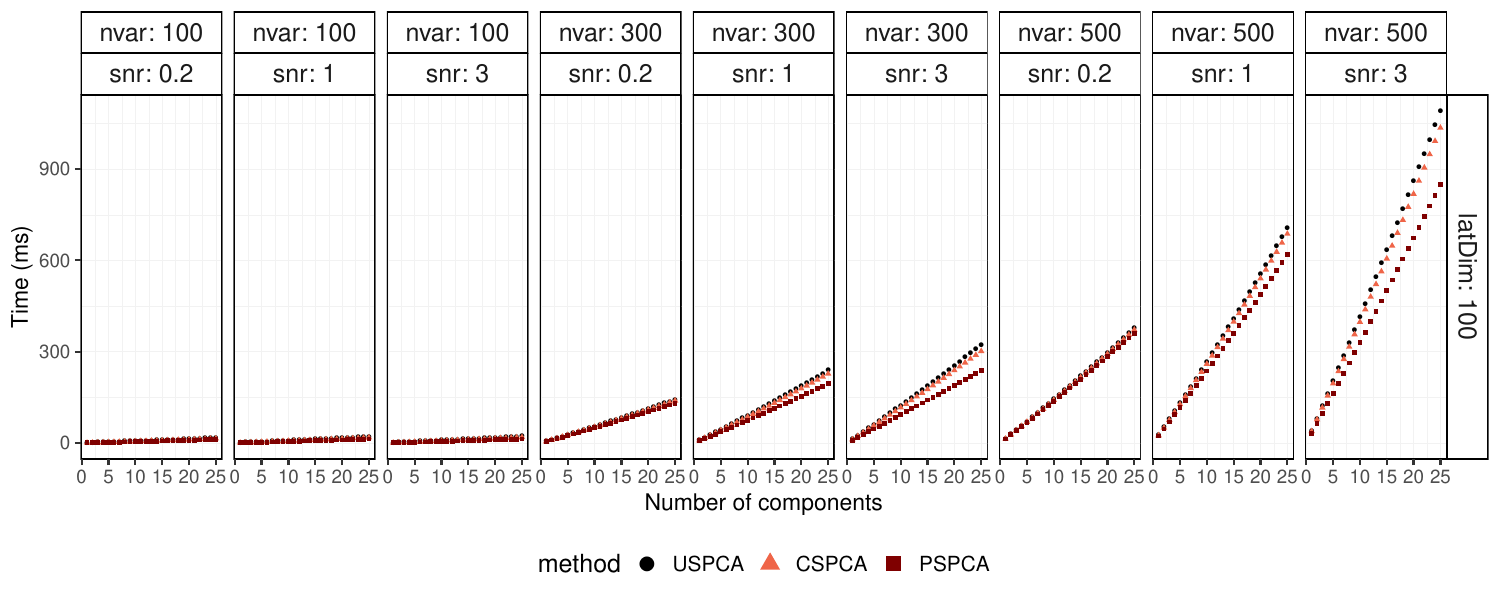}\\
  \caption{Median computational times (in milliseconds) for computing sparse components with underlying latent dimension of 100.}
  \label{fig:doetime}
\end{figure}

In order to assess the behavior of the different methods, USPCA, CSPCA and PSPCA, we simulated different datasets according to an experimental design. We considered three different levels of latent dimension (number of latent variables), $d \in \{5, 50, 100\}$; different number of variables, $p \in \{100, 300, 500\}$ and different signal to noise ratio (snr), $s \in \{0.2, 1, 3\}$. The model we considered is $\bX(d, p, s) = \bT\bP\trasp + \sqrt{s}\,\bE$, where $\bT$ are $d$ independent $\mathcal{N}(0, 1)$ latent variables, $\bP$ is a $p \times d$ matrix with unit-norm rows and $\bE$ is a matrix of $p$ independent $\mathcal{N}(0, 1)$ errors. Therefore, the theoretical correlation matrix of $\bX$ is equal to
$$
\bS = \mathrm{corr}(\bX) =  (\bP\bP\trasp + s\bI )/(1+s).
$$

When the snr, $s$, is small this correlation matrix is almost of rank $d$, while as snr increases the correlation matrix becomes closer to the identity matrix. The $\bP$ matrices were created by generating the entries as independent $\mathcal{U}(-1, 1)$ variables and rescaling the rows. The error correlation matrix was generated as the correlation matrix of a sample of $4p$ pseudo-random realizations of a $\mathcal{N}(0,1)$ variable, without removing possible random correlations. For each combination of levels we ran 1000 repetitions, computing five components when the latent dimension was equal to five and 25 when it was larger, for each method. Variables were selected using a forward selection with stopping criterion $\alpha = 0.95$. The PCs of the $\bQ_j$ matrices were computed with the power method but USPCA and CSPCA sparse loadings were computed with a direct generalized eigendecomposition algorithm.

In the following we highlight the main findings from these simulations. More details can be found in the Online Supplement to the paper.

The time taken to compute USPCA and CSPCA components are very similar and are indistinguishable on the plot. The higher efficiency of PSPCA shows when the number of variables or the snr grows, as can be appreciated by observing the median computational (CPU) times, shown in Figure~\ref{fig:doetime}.

\begin{table}[b!]
\centering
  \caption{Log-log regression of computational times on experimental factors.}

\bigskip
\begin{tabular}{Clllll}
          &       &       &       & \multicolumn{2}{c}{95\% Confidence Interval} \\
          \cmidrule{5-6}
Term  & \multicolumn{1}{l}{Estimate} & \multicolumn{1}{l}{Standard Error} & \multicolumn{1}{l}{$p$-value} & \multicolumn{1}{l}{Low} & \multicolumn{1}{l}{High} \\
\midrule
    Intercept  & \multicolumn{1}{r}{1.58} & \multicolumn{1}{r}{0.0050} & \multicolumn{1}{r}{0.0000} & \multicolumn{1}{r}{1.57} & \multicolumn{1}{r}{1.59} \\
    nvar (p)   & \multicolumn{1}{r}{2.21} & \multicolumn{1}{r}{0.0008} & \multicolumn{1}{r}{0.0000} & \multicolumn{1}{r}{2.21} & \multicolumn{1}{r}{2.21} \\
    latDim (d)  & \multicolumn{1}{r}{0.28} & \multicolumn{1}{r}{0.0018} & \multicolumn{1}{r}{0.0000} & \multicolumn{1}{r}{0.27} & \multicolumn{1}{r}{0.28} \\
    snr (s)   & \multicolumn{1}{r}{0.28} & \multicolumn{1}{r}{0.0005} & \multicolumn{1}{r}{0.0000} & \multicolumn{1}{r}{0.28} & \multicolumn{1}{r}{0.28} \\
    Comp. No. (j) & \multicolumn{1}{r}{1.17} & \multicolumn{1}{r}{0.0030} & \multicolumn{1}{r}{0.0000} & \multicolumn{1}{r}{1.16} & \multicolumn{1}{r}{1.17} \\
    CSPCA  & \multicolumn{1}{r}{$-0.04$} & \multicolumn{1}{r}{0.0012} & \multicolumn{1}{r}{0.0000} & \multicolumn{1}{r}{$-0.05$} & \multicolumn{1}{r}{-0.04} \\
    PSPCA  & \multicolumn{1}{r}{$-0.17$} & \multicolumn{1}{r}{0.0012} & \multicolumn{1}{r}{0.0000} & \multicolumn{1}{r}{$-0.17$} & \multicolumn{1}{r}{-0.17} \\
\bottomrule
      &       &       &       &       &  \\
\multicolumn{6}{l}{Residual standard error: 0.1446 on 80,993 degrees of freedom} \\
    \multicolumn{6}{l}{Multiple R$^2$:  0.9946,    Adjusted R$^2$:  0.9946 } \\
    \multicolumn{6}{l}{F-statistic:  2471982 on 6 and 80993 DF,  $p$-value: 0} \\
\end{tabular}
\label{tab:doeregtime}
\end{table}

We assumed a polynomial dependence of time on the parameters $d, p, s$ and the order of the component computed, $j$. Hence, we estimated the polynomial terms by regressing the logarithm of time on the logarithm of these parameters and adding indicator variables for the method used. The results of the regression, shown in Table~\ref{tab:doeregtime}, confirm the conclusions given above.
The fit is excellent, as indicated by the coefficient of determination $R^2 >0.99$, and the final time equation is
$$
t(d, p, s, j, M) = e^{1.58}	p^{2.21}	d^{0.28}	s^{0.28}	j^{1.17}	(0.96)^{I_{CSPCA}} ({0.84})^{I_{PSPCA}} \epsilon,
$$
where $I_M$ denotes the indicator variable equal to $1$ when method $M$ is used and $0$ otherwise. The coefficients of these indicator variables measure the ratio of time with the corresponding times taken by USPCA. This result confirms that using the power method to compute the PCs the complexity growth rate is about $O(p^{2.2})$. The time increases almost linearly with the number of components computed. PSPCA is slightly faster than the other methods.

The components computed to explain 95\% of the variance explained by the PCs have relatively low cardinality.  As expected, the cardinality of the components increases with the number of variables in the set, the latent dimension and the snr. The variability is low and it increases when the number of variables and the snr increase.

The results of the log-log regression of cardinality on the experimental factors gave an
excellent fit, with coefficient of determination $R^2 \approx 0.97$. The final cardinality equation is
$$
c(d, p, s, j, M) = e^{-0.73}	p^{0.49}	d^{0.48}	s^{0.39}	j^{1.01}	\epsilon.
$$
The cardinality increases less than linearly with the number of variables, latent dimension and snr, while it grows almost linearly with the components' order. The method used was not found to significantly change the cardinality of the solutions.

The proportions of variance explained by the three methods is very similar in value and in ratio, and differences are only observable at the third or fourth decimal figure. The variance explained by the CSPCA components is always very close to that explained by the USPCA components or is slightly higher. In most cases, the PSPCA components explain the least proportion of variance. The USPCA components of higher order tend to explain less variance than the CSPCA components. This phenomenon has already been observed and it is due to the greediness of the approach, when the local optimality of the USPCA components leads to globally inferior paths.

\begin{figure}[t!]
  \centering
  \includegraphics[width = 1\textwidth, height = 0.3\textwidth]{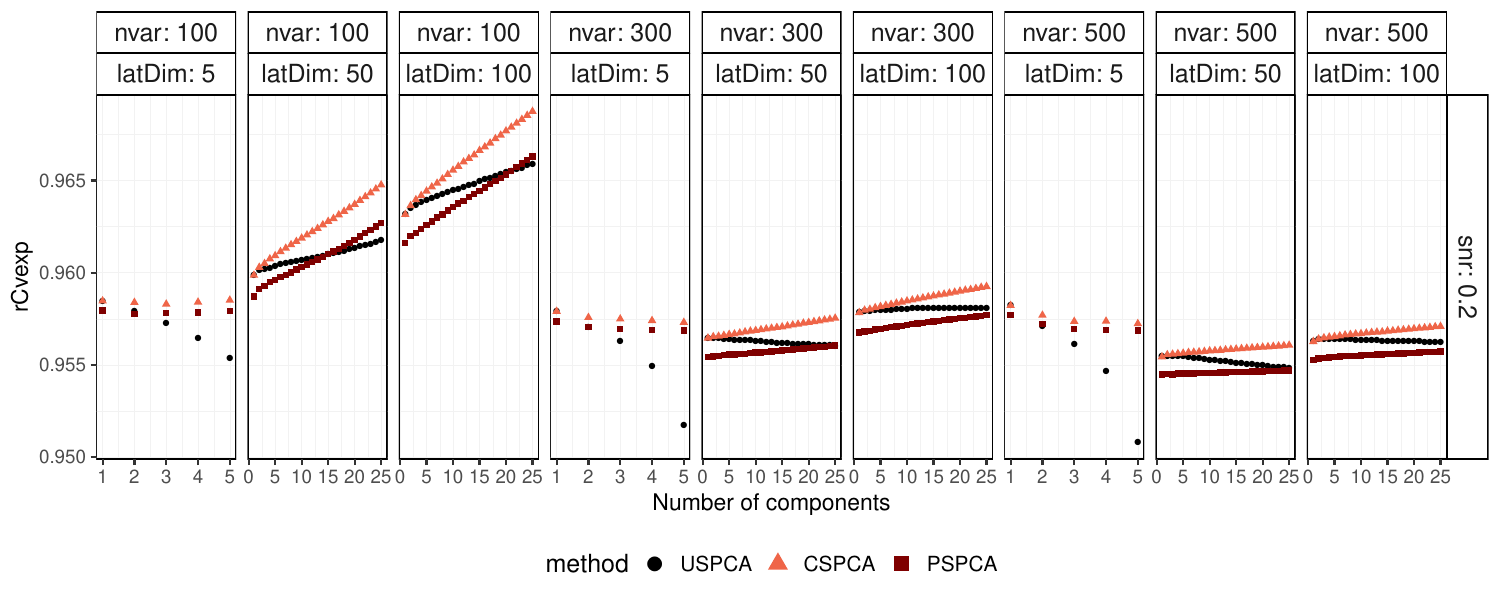}\\
  \caption{Median $\rCvexp$ for the sparse components computed with different methods for different simulated datasets, with constant snr = 0.2. The median values are computed over 1000 runs.}
  \label{fig:doecvexp}
\end{figure}

To compare the variance explained by different methods we use the cumulative variances explained by the sparse components relative to the variances explained by the same number of PCs,
$$
\rCvexp = {\sum_{i = 1}^j \evexp(\bt_i)}{\Big /} {\sum_{i = 1}^j\vexp(\bp_i)}.
$$
Figure~\ref{fig:doecvexp} shows the median $\rCvexp$ for various number of variables and latent dimensions at a constant snr, $s = 0.2$.
This shows how USPCA performs noticeably worse than the other methods when the latent dimension is small ($d = 5$) and the snr ratio is low. This is because, under this setting, the rank of the $\bQTj{Y}$ matrices is almost equal to $5 - j + 1$ and orthogonality determines a more severe departure from the optimal path. However, also in this case, the differences are very small.

Another aspect that we investigated is the correlation among the components computed with CSPCA and PSPCA. Since each component is highly correlated with the corresponding first PC of $\bQ_j$, which are mutually orthogonal, by Lemma \ref{le:corr}, we expect their correlation to be small. This is confirmed by the distribution of the $nc (nc-1)/2$ correlations between each pair of components computed for each experimental set up, of which the summary statistics are shown in Table~\ref{tab:corcomp}. The correlations are extremely small and do not show a particular pattern with respect to any of the experimental factors, except that, in most cases, the variability is slightly larger for the PSPCA components.

\begin{table}[b!]
  \centering
  \caption{Summary statistics of the correlations among sparse components computed with the same method on each simulated data set.}

      \bigskip
        \begin{tabular}{cccccc}
\toprule
     \multicolumn{1}{l}{Minimum} & \multicolumn{1}{l}{1st Quartile} & \multicolumn{1}{l}{Median} & \multicolumn{1}{l}{Mean} & \multicolumn{1}{l}{3rd Quartile} & \multicolumn{1}{l}{Maximum} \\
    0.004 & 0.005 & 0.006 & 0.008 & 0.009 & 0.016 \\
\bottomrule
\end{tabular}
  \label{tab:corcomp}
\end{table}

\subsection{Real datasets}

The datasets that we consider in this section, listed in Table~\ref{tab:descdata}, have been taken from various sources, mostly from the data distributed with the book ``Elements of Statistical Learning'' (ESL) \citep{has}. Other sets were taken from the UCI Machine Learning Repository \citep{uci}. The remaining sets were taken from different sources; see Table~\ref{tab:descdata} for details. Most of these are fat datasets  as they have a large number of features and fewer objects. The largest dataset, \cite{rama}, has been used to test other SPCA methods, including \citep{zou, sri, wan}.

\begin{table}[b!]
  \caption{Description of the datasets used for numerical comparison.}
 \centering

\begin{threeparttable}

\begin{tabular}{lrrlll}
Name  & Samples &Features & Type &Description & Source \\
\midrule
Crime& 1994& 99    &regular&  social data &UCI Repository\tnote{a} \\
Isolet & 6238& 716 &regular   & character recognition & UCI Repository\tnote{b} \\
Ross (NCI60) &60  & 1375 &fat & gene expression & \textsf{R} package \texttt{made4}\tnote{c}  \\
Khanh &88& 2308 &fat & gene expression & ESL\tnote{d} \\
Phoneme &257    & 4509 &fat & speech recognition & ESL \\
NCI60 &60   & 6830 &fat & gene expression & ESL \\
Protein &11    & 7466 &fat & protein cytometry& ESL \\
Radiation &58    & 12625 &fat & gene expression& ESL \\
Ramaswamy &144    & 16063 &fat & gene expression & Broadinstitute repository\tnote{e} \\
\bottomrule
\end{tabular}
\begin{tablenotes}
{\footnotesize \item[a] \url{https://archive.ics.uci.edu/ml/datasets/Communities+and+Crime}
\item[b] \url{https://archive.ics.uci.edu/ml/datasets/ISOLET}
\item[c] \url{http://bioconductor.org/packages/release/bioc/html/made4.html}
\item[d] \url{https://statweb.stanford.edu/~tibs/ElemStatLearn/}
\item[e] \url{http://software.broadinstitute.org/cancer/software/genepattern/datasets}}
\end{tablenotes}
\end{threeparttable}
\label{tab:descdata}
\end{table}

\begin{figure}[t!]
  \centering
  \includegraphics[width=0.90\textwidth, height = 0.45\textwidth]{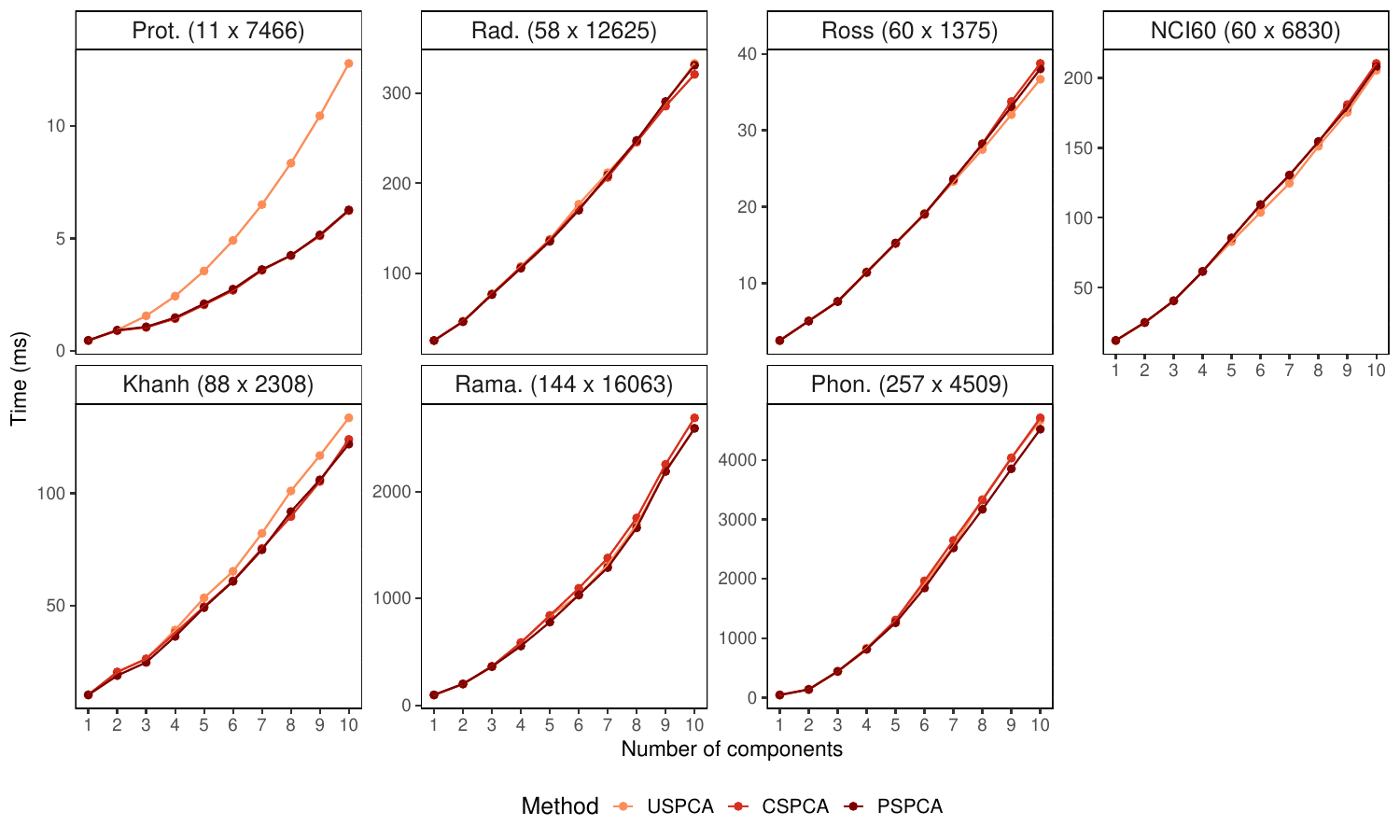}\\
   \caption{Median computational times taken to compute the first 10 sparse components with $\alpha = 0.95$ on
  seven fat datasets. Time is expressed in milliseconds.}
  \label{fig:fatty_times}
\end{figure}

First we compared the performance of USPCA, CSPCA and PSPCA on the fat dataset described in Table~\ref{tab:descdata}.
We computed 10 components for each dataset using the reverse svd approach and requiring that each of them explained at least a proportion $\alpha = 0.95$ of the variance explained by the corresponding PC. Both the PCs and the sparse loadings were computed using direct eigendecomposition.
Figure~\ref{fig:fatty_times} shows the median computational times over 25 repetitions. The plots are shown in increasing order of number of observations in the datasets.

\begin{figure}[H]
  \centering
  \includegraphics[width=1\textwidth, height = 0.5\textwidth]{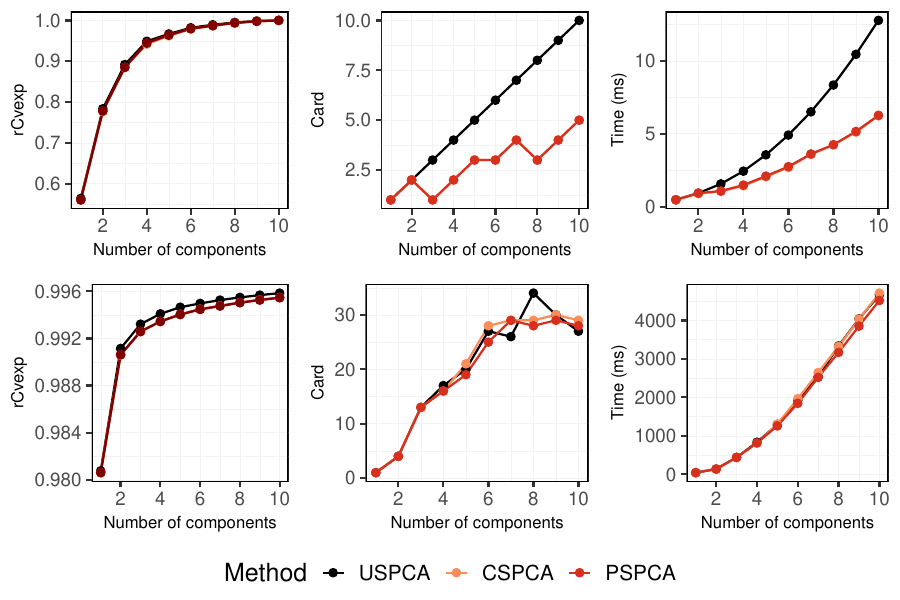}\\
     \caption{Comparison of the performance of LS SPCA methods on two fat datasets, Protein (top) and Phonema (bottom).}
  \label{fig:fatty_cards}
\end{figure}

The computational times of the three methods are very close for all datasets with the exception of Protein (Prot.). For this dataset the computation of the USPCA components takes longer because the orthogonality constraints require the cardinality to be larger than that of the other methods, as shown in Figure~\ref{fig:fatty_cards}.

Even though the computation of the eigenvectors of the $\bQj\bQ\trasps{j}$ matrices is $O(n^3)$, in some cases the computational time on some sets (for example Radiation) is greater than that on dataset with fewer variables and more observations (note the different scales on the vertical axes). This is because the computation of the PC loadings is $O(n^3 + n^2p)$, hence there is a cross-over effect, due to the number of variables.

\subsection{Comparison with conventional SPCA}\label{sec:compspca}

In this section we compare the performances of the first sparse components computed with a conventional SPCA method and with PSPCA. As our conventional SPCA method we used SPCA-IE \citep{wan} with the amvl criterion. This method was shown to perform similarly to other SPCA methods. It does not require to choose arbitrary sparsity parameters and is simple to implement.

Since the results for fat matrices are quite similar, we present the results only for four datasets: Crime, Isolets, Khanh and Ramaswamy. For the last dataset, we computed the conventional sparse components using simple thresholding, stopping the computation at cardinality 200; details of the performance of components with larger cardinality computed with different conventional SPCA methods for this dataset can be found in the papers cited above.

\begin{figure}[t!]
  \centering
  \includegraphics[width=0.95\textwidth, height = 0.5\textwidth]{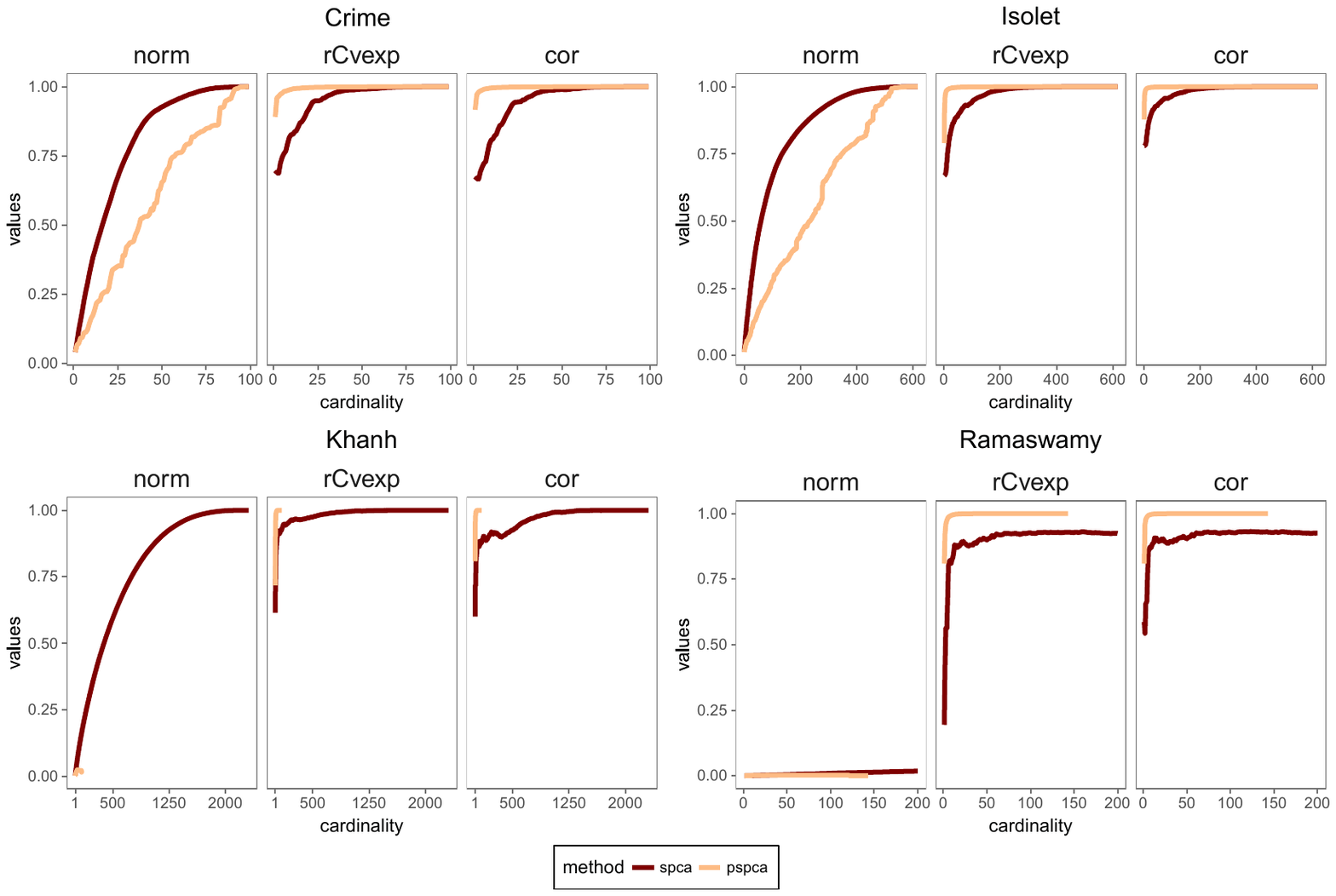}\\ 
  \caption{Norm, $\rCvexp$ and correlation with the PCs versus cardinality of the first sparse components computed with SPCA-IE and PSPCA.}
  \label{fig:compspca}
\end{figure}

Figure~\ref{fig:compspca} compares the relative norm ($||\bt||^2/||\bX||^2$), $\rCvexp$ and their correlation with the full PCs for increasing cardinality of the first components computed with PSPCA and SPCA-IE on different datasets. The PSPCA values for the rank deficient Khanh and Ramaswamy datasets are available only until the solutions reach full rank cardinality (87 and 143, respectively) at which the components explain the maximum possible variance. Clearly SPCA-IE outperforms PSPCA in the norm of the components. However, the latter method guarantees higher variance explained and closer convergence to the PC with much lower cardinality.

The differences in performance of the two approaches are more evident for large rank deficient datasets, when conventional sparse components with cardinality in the hundreds explain less variance than PSPCA components of much lower cardinality. The plots also show clearly that the components' norms are not related to the variance that they explain or to their correlation with the PC. This confirms the theoretical conclusions given in Section \ref{sec:convspca}.

\begin{table}[b!]
 \centering
  \caption{Cardinality needed to reach 99.9\% $\rCvexp$ by the components computed with PSPCA.}
\begin{threeparttable}

\bigskip
\begin{tabular}{lrrr}
 Dataset& Rank     & \multicolumn{2}{c}{Cardinality Needed for $99.9\%\, \rCvexp$}\\
\cmidrule{3-4}
&& PSPCA & SPCA\tnote{a}\\
\midrule
Crime & 99& 38    & 74    \\
Isolet & 716& 123   & 439 \\
Khanh & 87& 28    & 1338  \\
Ramaswamy &143&  23    & $>$ 200\\
\bottomrule
\end{tabular}
\begin{tablenotes}
\item[a] The values for the first three datasets were obtained using SPCA-IE and the values for the Ramaswamy dataset using simple thresholding
\end{tablenotes}
\end{threeparttable}
\label{tab:compspca}%
\end{table}

Table~\ref{tab:compspca} shows the cardinality with which the components computed with the two methods reached 99.9\% $\rCvexp$. In all cases the cardinality of the PSPCA components is much lower than that of the SPCA-IE components.

\section{Discussion}\label{sec:conclusions}

Projection SPCA is a very efficient method for selecting variables for computing a sparse approximation to the PCs. The methodology is intuitive and can be understood by users who do not have a deep knowledge of numerical optimization. The only parameter to be set for computing the solutions is the proportion of variance explained, the meaning of which is also easily understandable. The algorithm is simple to implement and scalable to large datasets. Users can choose their preferred regression variable selection algorithm to select the variables. Most conventional SPCA methods, instead, are based on special numerical optimizations methods and require setting values for parameters with a difficult to understand effect. Future research could explore the results of using $\ell_1$ norm selection methods, such as least angle or Lasso regression, for example, on the computation of LS SPCA components.

In this work we have developed a framework for computing LS SPCA components that closely approximate the full PCs with low cardinality. We showed that sparse USPCA, CSPCA and PSPCA components can be efficiently computed for very large datasets. We also show that conventional SPCA methods suffer from a number of drawbacks which yield less attractive solutions than the corresponding LS SPCA solutions.

Conventional SPCA methods have been shown to give results similar to simple thresholding if not worse; see, e.g., \citep{zou, wan}. Thresholding has been proven to give misleading results; see, e.g., \citep{cad}. Since the loadings are proportional to the covariances of the variables with the PCs, the largest loadings correspond to variables that are highly correlated with the current PC and among themselves. These sets of variables are not very informative because they contain different measures of the same features.

\citet{zou} proposed three properties of a good SPCA method:
\begin{itemize}
\item [(i)]
Without any sparsity constraint, the method should reduce to PCA.
\item [(ii)]
It should be computationally efficient for both small \emph{p} and big \emph{p} data.
\item [(iii)]
It should avoid misidentifying the important variables.
\end{itemize}
The first property is not enough for a good method. The second is not necessary for the most commonly analyzed datasets and the third is vague because importance is not defined and variables known to be unimportant could be directly eliminated from the analysis.

We suggest the following properties for a good SPCA method:
\begin{itemize}
\item [(i)]
Without any sparsity constraint, the method should reduce to PCA.
\item [(ii)]
It should identify the sparsest expression of the principal components.
\item [(iii)]
The addition of a variable perfectly correlated with one or more variables already in the solution should not improve the objective function.
\end{itemize}
The last property eliminates redundant variables from the solution and should deter the inclusion of highly correlated ones. Conventional SPCA methods do not have the last two properties while methods based on LS SPCA do. It is possible that other methods could have these properties.

LS SPCA is implemented in the \textsf{R} package \texttt{spca} available on GitHub. PSPCA will be added to this package in the future.

\section*{Acknowledgments}

We thank the Editor-in-Chief, Christian Genest, an Associate Editor and the anonymous referees for their useful comments and suggestions that improved the paper. Gemai Chen's research is partially supported by a grant from the Natural Sciences and Engineering Research Council of Canada.

\section*{Appendix}

\setcounter{equation}{0}
\setcounter{proposition}{0}
\renewcommand{\theequation}{A.\arabic{equation}}
\renewcommand{\theproposition}{\Alph{proposition}}
\renewcommand{\thetheorem}{\Alph{theorem}}
\renewcommand{\thelemma}{\Alph{lemma}}

\begin{proposition}
Given an ordered set of $d$ components, $\bt_j = \bX\ba_j$ with $ j \in \{ 1, \ldots, d\}$,
the different types of variance explained as defined in Eqs.~\eqref{eq:vexp1}, \eqref{eq:evexp} and \eqref{eq:vexpq} satisfy
\begin{equation*}
\vexp_Q(\btj) \leq \evexp(\btj) \leq \vexp(\btj),
\end{equation*}
where $\vexp(\btj) =  {\btjT\bX\bXT\btj}/{\btjT\btj}$.
Equality is achieved for the first component or if a component is orthogonal to the preceding ones.
\end{proposition}

\medskip
\noindent
\textbf{Proof}.
The extra variance explained cannot be smaller than the variance of $\bQ$ explained by the same components because
\begin{equation*}
\vexp_Q(\btj) =  \evexp(\bt_j) \, \frac{\bajT\bQTjT{T}\bQTj{T}\baj}{\bajT\bXT\bX\baj}
 = \evexp(\bt_j) \left(\frac{\bajT\bXT\bX\baj - \bajT\bXT\Pi_{\bTjmu}\bX\baj}{\bajT\bXT\bX\baj} \right) \leq
\evexp(\bt_j).
\end{equation*}
It is well known that extra variance explained is not larger than the variance explained by a regressor. In fact,
\begin{equation*}
\vexp(\btj) = ||\Pi_{\btj}\bX||^2 = ||\Pi_{[\Pi_{\bT_{[j-1]}}\btj]}\bX + \Pi_{[\bQTj{T}\baj]}\bX||^2 \geq
||\Pi_{[\bQTj{T}\baj]}\bX||^2 = \evexp(\btj)
\end{equation*}
because $\bt_j = \Pi_{\bT_{[j-1]}}\btj + \Pi_{[(\bI - \bT_{[j-1]})]}\btj$ and
$\Pi_{[(\bI - \bT_{[j-1]})\btj]} = \Pi_{[\bQTj{T}\baj]}$. Therefore, $\Pi_{\btj} = \Pi_{\Pi_{\bT_{[j-1]}}\btj} + \Pi_{\Pi_{(\bI - \bT_{[j-1]})\btj}}$; see, e.g., Theorem~8.8 in~\citep{pun}.

The statement about the equality is true because if a component $\btj$ is orthogonal to all preceding variables, then $\btj = \bX\ba_j = \bQTj{T}\baj$, and $\bQ_{\bT_{[0]}}= \bX$.\hfill $\Box$

\setcounter{theorem}{0}
\begin{theorem}
Let $\dXj$ be a block of linearly independent variables. Then,
\begin{enumerate}
\item[(i)] The orthogonal LS SPCA components, $\by_j = \dXj\dbj$, are the first PCs of the matrices $(\Pi_{\dXj} - \Pi_{\hat{Y}_\dXj} )\bX$, where $\hat{Y}_\dXj = \Pi_{\dXj}\bY_{[j-1]}$.
\item[(ii)]
The nonorthogonal LS SPCA components, $\bzj = \dXj\ddj$, are the first PCs of the matrices $\hQTj{Z} = \Pi_{\dXj}\bQTj{Z} = \Pi_{\dXj} (\bI - \Pi_{\bZ_{[j-1]}} )\bX$.
\end{enumerate}
\end{theorem}

\medskip
\noindent
\textbf{Proof}.
Premultiplying Eq.~(\ref{eq:uspca}) by $\dXj\dSjinv$ gives
\begin{equation*}
  \Big(\Pi_{\dXj} - \Pi_{\hat{Y}_\dXj}\Big)\bX\bX\trasp \by_j = \by_j\xi_{j_\text{max}},
\end{equation*}
where $\Pi_{\hat{Y}_\dXj} = \Pi_{\dXj}
\bYjmu  (\bYjmuT\Pi_{\dXj}\bYjmu )\matinv\bYjmuT\Pi_{\dXj}$.
Since, $ (\Pi_{\dXj} - \Pi_{\hat{Y}_\dXj} )\by_j = \by_j$, we can write
$$
\Big(\Pi_{\dXj} - \Pi_{\hat{Y}_\dXj}\Big) \bX\bX\trasp \Big(\Pi_{\dXj} - \Pi_{\hat{Y}_\dXj}\Big)\by_j =
\by_j\xi_{j_\text{max}},
$$
which proves part (i). Given that $ \cspace(\hat{Y}_\dXj)\subset \cspace(\dXj)$, $\Pi_{\dXj} - \Pi_{\hat{Y}_\dXj} $ is a projector onto $\cspace(\dXj)\cap \cspace(\hat{Y}_\dXj)^\bot$, where $\cspace(\bA)^\bot$ denotes the orthocomplement of $\cspace(\bA)$ with respect to $\bI$; see Chapter~7 in~\citep{pun}.

In a similar fashion, premultiplying Eq.~\eqref{eq:lsspcasol} by $\dXj\bdS\matinvj{j}$ we obtain
$$
\Pi_{\dXj}\bQTj{Z}\bQTjT{Z}\bzj = \hQTj{Z}\hQTjT{Z}\bzj = \bzj\gamma_j,
$$
which proves part~(ii).\hfill $\Box$

\setcounter{lemma}{0}
\begin{lemma}
Let $t$ and $x$ be two random variables and $\by = (y_1, \ldots, y_d) \trasp$ a set of $d$ random variables uncorrelated with $x$.
If $\mathrm{corr}^2(t, x) =\alpha$, then for all $i \in \{ 1, \ldots, d\}$,
$   \mathrm{corr}^2(t, y_i) \leq 1 - \alpha$.
    If the $y_i$ variables are mutually uncorrelated, it follows that
    \begin{equation*}
    \sum_{i=1}^d \mathrm{corr}^2(t, y_i) \leq 1 - \alpha.
     \end{equation*}
\end{lemma}

\medskip
\noindent
\textbf{Proof.}
Let  $\boldsymbol{\rho}\trasps{t\by} = ( \rho_{ty_1}, \ldots,\rho_{ty_d})$, where $\rho_{ty_i}= \mathrm{corr}(t,y_i)$. The squared multiple correlation coefficient of the regression of $t$ on $[x,\mathbf{y}\trasp]\trasp$ is such that
\[
\rho_{t.x\by}^2 = [\sqrt{\alpha}, \boldsymbol{\rho}\trasps{t\by}]
    \begin{bmatrix}
        1& \mathbf{0}\trasp\\
        \mathbf{0} & \bR\matinv
        \end{bmatrix}
        \begin{bmatrix}
        \sqrt{\alpha}\\ \boldsymbol{\rho}_{t\by}
    \end{bmatrix}
    = \alpha + \rho_{t.\by}^2\leq 1,
\]
where $\bR$ is the correlation matrix of $\by$ and $\rho_{t.\by}^2  = \boldsymbol{\rho}\trasps{t\by} \bR\matinv\boldsymbol{\rho}_{t\by}$ is the squared coefficient of multiple correlation between $t$ and $\by$.

Since $\rho_{t.\by}^2$ can be written as the sum of the squared correlation of the response variable with one of the regressors, $y_i$, say, and the multiple correlation of the response variable with the orthogonal complement of the remaining variables, $\{y_j,\ j\neq i\}$, namely, $\rho_{t.\by}^2 = \mathrm{corr}^2(t,y_i)  + \rho_{t.\by_{/i}.y_i}^2\geq \mathrm{corr}^2(t,y_i)$, it follows that
\[
 0\leq \mathrm{corr}^2(t,y_i) \leq \boldsymbol{\rho}\trasps{t\by} \bR\matinv\mathbf{\rho}_{t\by} \leq 1 - \alpha.
\]
When $\mathrm{corr}(y_i,y_j)=0, i\neq j$,
$$
\rho_{t.\by}^2 = \boldsymbol{\rho}\trasps{t\by} \bR\matinv\boldsymbol{\rho}_{t\by} =
\sum_{i=1}^p \mathrm{corr}^2(\bt, \by_i),
$$
from which follows the second statement to be proved. For a more general proof, see \cite{pun05}.\hfill $\Box$

\begin{lemma}
When $\mathrm{rank}(\bX) = r < p$ the principal components can be expressed as sparse components of cardinality $r$ and loadings that have norm larger than $1$.
\end{lemma}

\medskip
\noindent
\textbf{Proof}.
Given that $\mathrm{rank}(\bX) = r$, there are $p - r$ columns of $\bX$ which are linearly dependent. Assume without loss of generality that the first $r$ columns of $\bX$ are linearly independent and denote them as $\dX$.

Also let $\cX$ be the remaining columns. Then, we can write
\begin{equation*}
\bX = [\dX, \cX] = \dX \Big\{ \bI_r, \dX\trasp(\dX\dX\trasp)\matginv\cX \Big\} =  \dX \Big \{\bI_r, (\dX\trasp\dX)\matinv\dX\trasp\cX\Big\} \nonumber
=\Pi_{\dX}\bX = \dX\bG,
\end{equation*}
where the superscript `$^+$' denotes the Moore--Penrose generalized inverse and $\bG = \bI_r, (\dX\trasp\dX)\matinv \dX\trasp\cX$.

Hence, the PCs can be written as $\bu_j = \bX\bv_j = \dX(\bG\bv_j) = \dX\dot\bv_j$, where $\dot\bv_j = \bG\bv_j$ has length $r$. Since the largest singular value of $\bX_j$ ($\bQTj{U}$) is not smaller than the largest singular value of $\dXj$ ($\dQTj{U}$), it must follow that $\dot\bv\trasps{j}\dot\bv_j \geq 1$. Therefore, the PCs can be defined as sparse components of cardinality $r$ with loadings of norm larger than $1$.\hfill $\Box$

\bigskip
\begin{state}
The CSPCA loadings can be computed as the generalized eigenvectors satisfying
$$\bC_j \bJ\trasps{j}\bS\bS\bJ_j \bC\trasps{j}\dbj = \dSj\dbj\xi_{j},
$$
where $\bC_j = \bI_{c_j} -  \bHjT(\bHj\bdS\matinvj{j}\bHjT)\matinv\bHj\bdS\matinvj{j}$.
\end{state}

\medskip
\noindent
\textbf{Proof}.
From Proposition~\ref{prop:uspca}, we have that the USPCA loadings satisfy
\begin{equation*}
\bC_j\dXjT\bX\bXT\dXj\dbj = \bC_j\bJ\trasps{j}\bS\bS\bJ_j\dbj = \dSj\dbj\xi_{j}.
\end{equation*}
Then
$$
    \bC_j\bJ\trasps{j}\bS\bS\bJ_j\dbj = \dSj\dbj\xi_{j} \quad \Leftrightarrow \quad
    \bC_j\bJ\trasps{j}\bS\bS\bJ_j \dSjinv (\dSj\dbj ) = \dSj\dbj\xi_{j}.
$$
Since $\bC_j$ is idempotent and  $\cspace(\dSj\dbj)\subseteq\cspace(\bC_j)$, because $\bC_j\bJ\trasps{j}\bS\bS\bJ_j
  \dSjinv (\dSj\dbj ) \propto \dSj\dbj$, $\dbj$ must satisfy
$$  \bC_j\bJ\trasps{j}\bS\bS\bJ_j
  \dSjinv\bC_j (\dSj\dbj ) = \bC_j\bJ\trasps{j}\bS\bS\bJ_j (\dSjinv\bC_j\dSj )\dbj =
  \dSj\dbj\xi_{j}.
$$
Now,
\begin{equation*}
\dSjinv\bC_j\dSj = \dSjinv \Big \{ \bI - \bHjT (\bHj\dSjinv\bHjT )\matinv\bHj\dSjinv \Big \} \dSj =
\bI - \dSjinv\bHjT (\bHj\dSjinv\bHjT )\matinv\bHj = \bC\trasps{j}.
\end{equation*}
Therefore, $\dbj$ is the generalized eigenvector satisfying $ \bC_j\bJ\trasps{j}\bS\bS\bJ_j
 \bC\trasps{j}\dbj = \dSj\dbj\xi_{j}$. This completes the argument.\hfill $\Box$

\end{document}